\documentclass[iop,apj]{emulateapj}
\usepackage{amsmath, amsthm, amssymb, amsfonts}
\usepackage{graphicx}
\usepackage{dcolumn}
\usepackage{natbib}
\usepackage{bm}
\usepackage[caption=false]{subfig}
\usepackage{verbatim}

\begin{document}

\title{The Polarization Behavior of Relativistic Synchrotron Jets}

\author{A. L. Peirson\altaffilmark{1} \& Roger W. Romani\altaffilmark{1}}
\affil{\altaffilmark{1}Dept. of Physics and Kavli Institute for Particle Astrophysics and Cosmology, Stanford University, Stanford, CA 94305}
    
\begin{abstract}
\vspace{5pt}
\normalsize{We describe a geometric model for synchrotron radiation from blazar jets, involving multiple emission zones with turbulent magnetic fields and a transient core with a helical B field. Including the effects of jet divergence, particle cooling and the Relativistic PA rotation (RPAR) to the observer frame, we find polarization behavior consistent with recent data from monitoring campaigns. We predict that under some circumstances multi-$\pi$ rotation phases should exhibit relativistically-induced steps in rate $d{\rm PA}/dt$ and modulation in polarization $\Pi$ that can be helpful in pinning down the jet $\Gamma$ and $\theta_{\rm obs}$. Also, RPAR enhances waveband differences that will be particularly interesting for comparing radio, optical and, soon, X-ray PA and $\Pi$ variations.
}
\end{abstract}

\pacs{Valid PACS appear here}
\maketitle


\section{\label{sec:level1}Introduction}
Blazars are active galactic nuclei whose powerful relativistic jets point at small angle $\theta_{\rm obs}$  to the Earth line-of-sight \citep{urry_unified_1995}, so that the Doppler-boosted jet emission dominates the observed spectral energy distribution (SED). This SED is characterized by a low energy peak caused by synchrotron radiation from energetic electrons, and a high energy peak generally attributed to Inverse Compton scattering of photons by these same electrons \citep{maraschi_jet_1992}. The seed photons can either be from the synchrotron emission (SSC) or from an external source such as the accretion disk or broad line region (EC). The sources are further subdivided by the frequency of the $\nu F_\nu$ synchrotron peak \citep{abdo_spectral_2010}, with ${\rm log}\,\nu_{\rm sy} <14$ labeled LBL (Low peak BL Lacs, and most Flat spectrum Radio Quasars FSRQ) and ${\rm log}\,\nu_{\rm sy} > 15$ called HBL (High peak BL Lacs). Here the frequency is in Hz, and IBL represent the intermediate case. We have yet to determine how the jets are energized and launched with bulk Lorentz factor $\Gamma$, but an attractive origin is the \citet{blandford_electromagnetic_1977} process, so that the jet axis may be associated with the spin axis of the central black hole and the angular momentum axis of the surrounding accretion disk. The jet $e^+/e^-$ obtain an energy distribution extending to $\gamma_{\rm max} \sim 10^4$ or higher, often attributed to shock acceleration. Radiation from these particles spiraling in the embedded magnetic field $B$ can be used to constrain the geometry and energetics of the emission zone and, by inference, the jet accelerator.

In studying jet geometry polarization can be particularly useful. Radio VLBI studies have long shown that the pc-scale jet can be substantially polarized. Recently much effort has been spent on measuring the optical polarization properties of blazars, since this probes even smaller scales, closer to the acceleration zone. This polarization is often quite variable, offering new dynamical information on the jet structure \citep[e.g.][]{blinov_robopol:_2015, lynch_green_2018}.
In the near future we also hope to measure the X-ray polarization of a significant population of blazars with {\it IXPE} \citep{weisskopf_imaging_2016} and similar new facilities. 
\subsection{Blazar EVPA Variability}

The polarization fraction $\Pi$ and electric vector position angle $\theta_{\rm EVPA}$ of blazar emission have long been known to exhibit stochastic variability. Indeed optical polarization variability is a defining property of the BL Lac class. Recent monitoring campaigns have revealed new polarization patterns. The typical behavior is a stochastic variation about $\Pi\sim 0.05-0.15$ fluctuating with $\Pi/\sigma_\Pi$ and $\sigma_{\theta_{\rm EVPA}} \sim 1$. In addition, periods of relatively steady rotation of the EVPA, sometimes extending many $\times \pi$, can occur lasting weeks or months \citep{blinov_robopol:_2015}, after which the EVPA returns to the stochastic phase. These may be associated with flares in the total intensity \citep{blinov_robopol:_2016}, but this is not always the case.

	A few other trends have been noted. \cite{blinov_robopol:_2016} indicate that $\Pi$ is on average smaller in the rotating phases. However there are many examples where $\pi$ increases during rotation; these seem more common for the long $\Delta {\rm PA} > \pi$ rotations (I. Liodakis, priv. comm.). Also $\sigma_\Pi$ of a given source appears to be similar in the stochastic and rotating phases. Thus while a systematic B structure should be present to drive the large angle EVPA swings an underlying stochastic process must continue. Further, there is a tendency for the mean EVPA during the stochastic phase to correlate on the sky with the projected jet axis \citep{jorstad_multifrequency_2006}. Radio and optical polarization behavior can be similar \citep{darcangelo_synchronous_2009}, with $\Pi$ often higher in the optical. There appear to be an association between GeV flares and rotation phases \citep{blinov_robopol:_2018}. Also rotations appear to be recurrent in some blazars and absent in others. However none of these trends is universal. An example of a source making a transition from stochastic to rotation phase and back is shown in Figure 1.
\\  
\begin{figure}[t]

\includegraphics[width=0.9\linewidth, height=6.0cm]{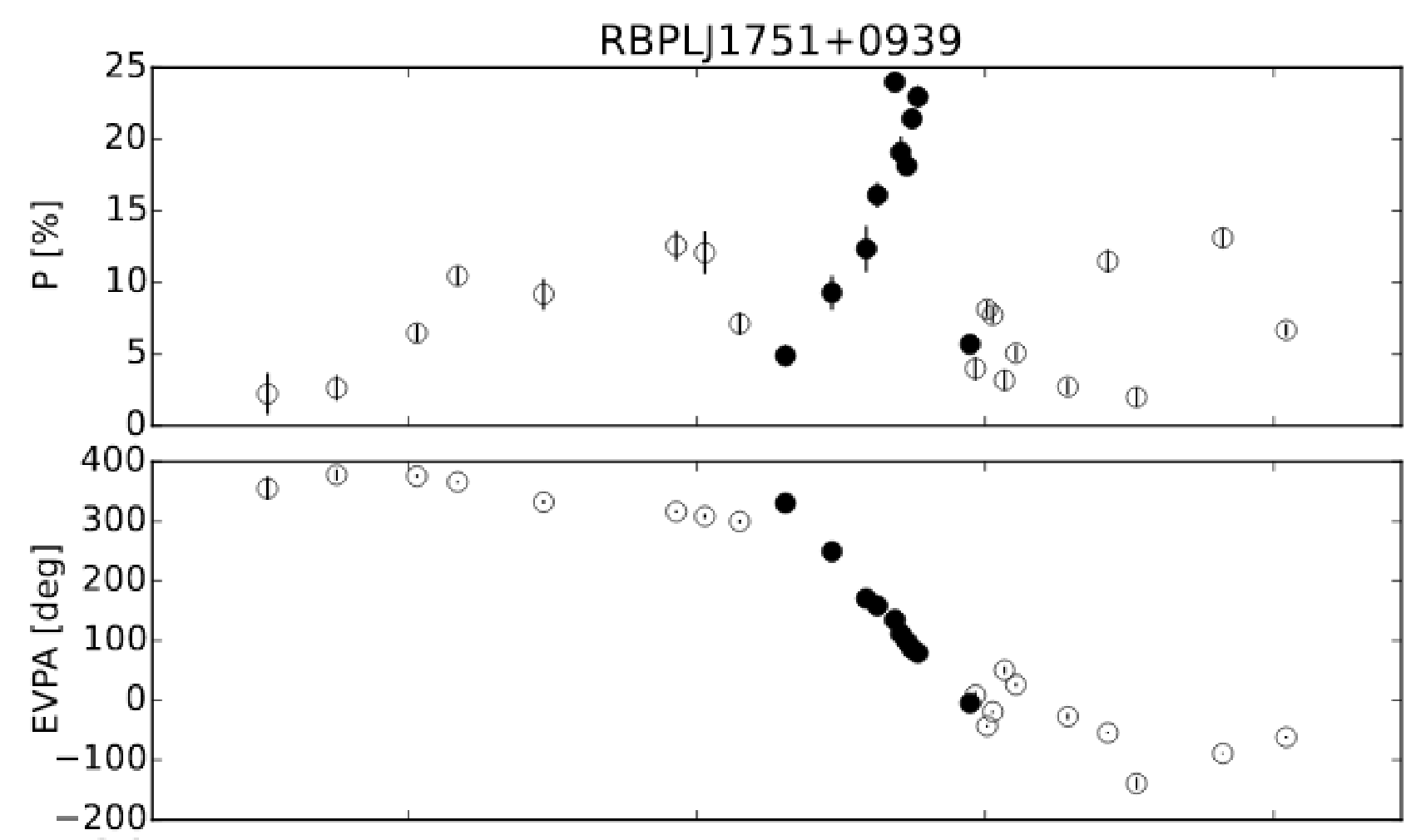}
\centering
\caption{The polarization fraction and EVPA in the R-band against the time in days during an observed rotation by RoboPol \cite{blinov_robopol:_2016} of the LBL blazar J1751+0939. Filled points mark their identification of a `rotating' epoch.}
\end{figure}

	Attempts to model such behavior have taken a variety of forms. In \cite{hughes_synchrotron_1989} a set of multiple shocks in the jet was used to reproduce the stochastic fluctuations, while in \citet{marscher_turbulent_2014} multizone turbulence in a conical standing shock was posited to generate the stochastically variable emission. Such stochastic variation can induce epochs of relatively constant PA sweep, but it was shown \citep{blinov_robopol:_2016} that the incidence (and persistence over many $\times \pi$) of the observed sweeps are inconsistent with purely stochastic models. Accordingly, models for the rotating phase generally invoke helical structures in the jet. For example \citet{zhang_polarization_2015} invoke a helical field energized by a standing shock. Such fields are suggested by Faraday rotation gradients transverse to the local jet directions in several nearby blazars e.g. PKS 0745+241, PKS 0820+225, Mrk 501, 3C 371 \citep{gabuzda_helical_2004} and could quite naturally be attributed to field symmetries imposed at the jet base by the Blandford-Znajek process \citep{blandford_electromagnetic_1977}. Alternatively, \citet{lyutikov_polarization_2017} assume that the jet itself takes on a helical form, e.g. due to precession, with an aligned embedded field. A third picture \citep{nalewajko_model_2017} posits a helical kink propagating along a conical jet with an embedded toroidal B field. Each of these pictures can accommodate smooth multicycle rotations.

	We explore here a heuristic model that incorporates the main features above, in an attempt to reproduce the range of observed optical polarization phenomena and to predict new correlations to be tested with multiwavelength observations. We start (\textsection2) with a description of the important, but under-appreciated, effect of relativistic boosting on the observed polarization. We then describe (\textsection3) a toy geometry with multiple zones transitioning between random and helical magnetic patterns and propagating downstream in a conical jet. We then couple this with a radiation model that follows the cooling and synchrotron radiation of the $e^+/e^-$ (\textsection4) and comment on the patterns and multiwavelength correlations of the resulting polarization signal. \textsection5 applies this picture to the bright HBL Mrk 501, and we conclude with general predictions for future extensions and comparisons with the data.

\section{\label{sec:level1}Relativistic PA Rotation}
\begin{figure*}[ht!]
  \centering
  \includegraphics[width=16cm,height=8cm]{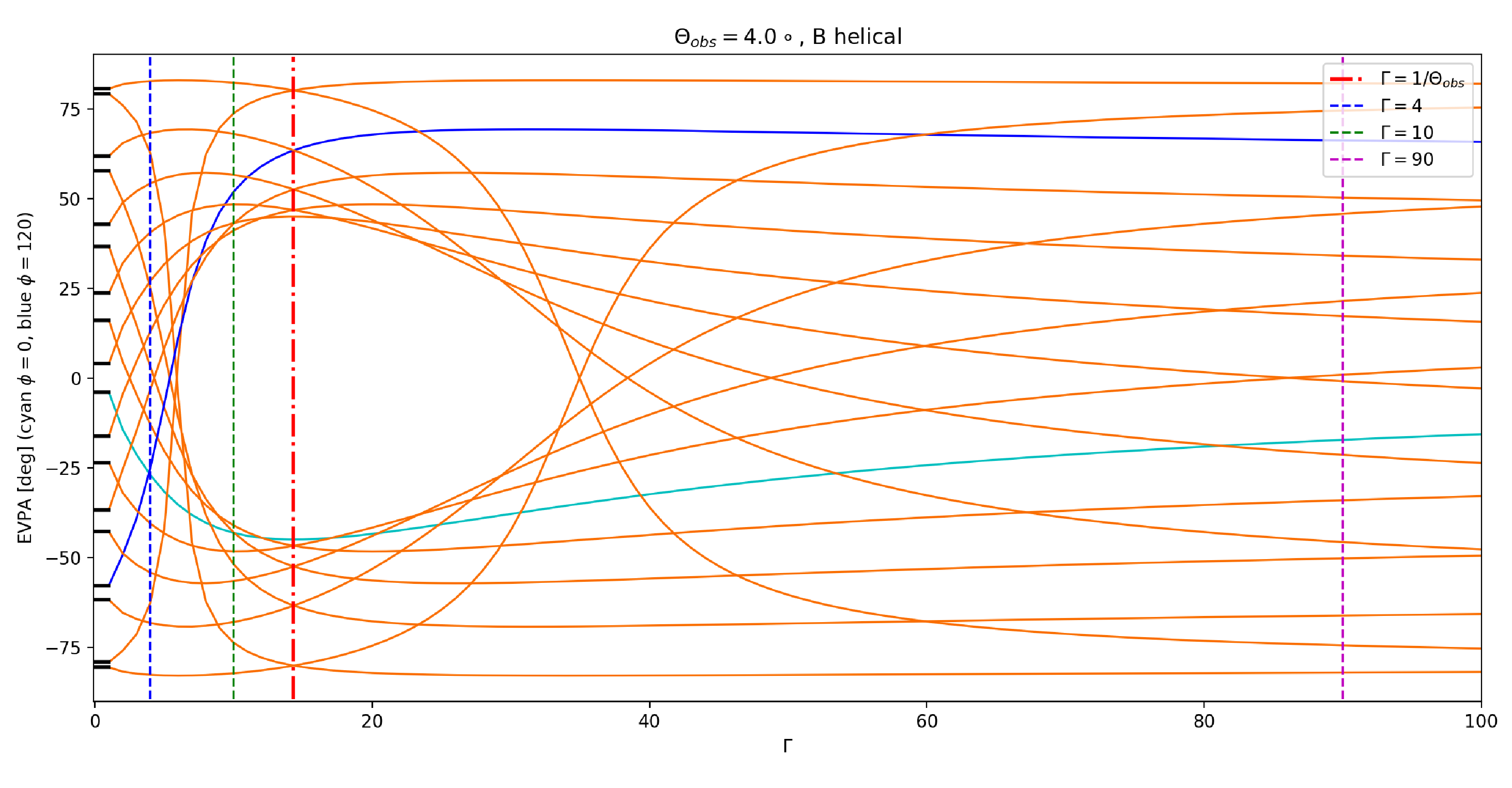}
  \caption{EVPA in the observers frame on the plane of the sky as a function of jet bulk gamma $\Gamma$ for a helical B-field with a 45$^{\circ}$ pitch angle. The direction of the jet is 4$^{\circ}$ off our line of sight in the $\hat{x}$ direction on the plane of the sky, corresponding to 0$^{\circ}$ on the plot. The B-field is sampled every $\phi_B =20^{\circ}$ from $0 - 360^{\circ}$. The black lines on the y-axis denote the observed EVPAs without RPAR (i.e. $\Gamma = 1$). The solid cyan and blue lines mark $\phi_B =0^{\circ},120^{\circ}$ respectively.}
\end{figure*}
\begin{figure*}[ht!]
  \centering
  \includegraphics[width=16cm,height=8cm]{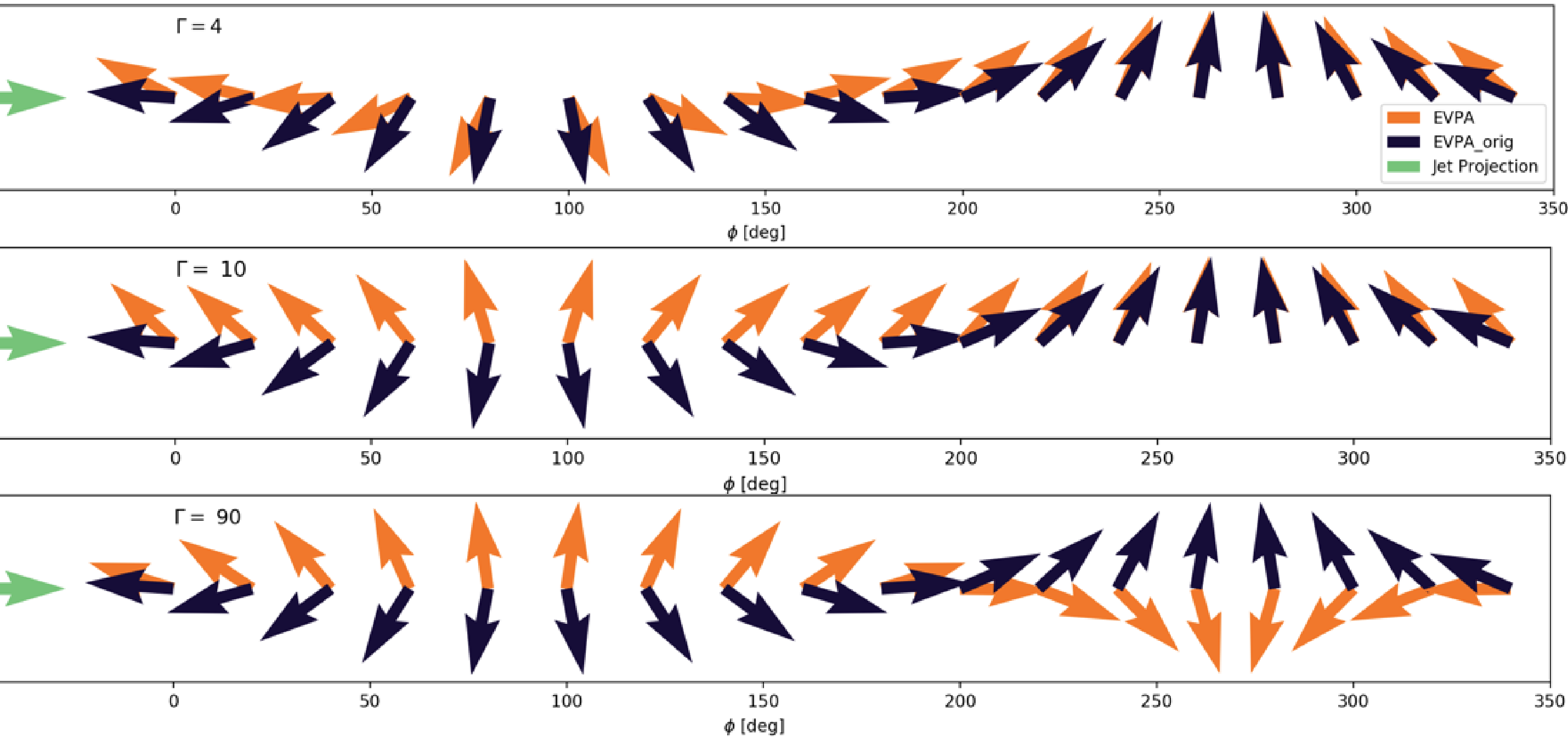}
  \caption{Arrows denoting the observed EVPA on the plane of the sky for the jet described in Figure 2 at the 3 different $\Gamma$ cuts. The black arrows show the initial EVPA absent RPAR (i.e. for stationary $\Gamma = 1$), the orange arrows show the RPAR-shifted EVPAs. The projected jet axis is along the green arrow. Arrows are plotted as single headed for clarity, although observations will span 180$^\circ$ (i.e. -90 to +90 in Figure 2).}
\end{figure*}

	In most cases models assume that the observed polarization direction is that of the EVPA in the jet fluid frame $\theta_{\rm jf}$. However, \citet{lyutikov_polarization_2003}, following  \citet{blandford_relativistic_1979}, show that the relativistic transformations can induce significant rotation, with the direction and magnitude of the observed angle depending on $\theta_{\rm jf}$, $\Gamma_{\rm bulk}$, $\theta_{\rm obs}$ and $\underline{\hat{v}}$ (the direction of the jet). We refer to this process as Relativistic PA rotation (RPAR) in the following discussion. Taking $\mathbf{\hat{e}' = \hat{n}' \times \hat{B}'}$ as the EVPA in the jet frame, they show that 
\begin{equation}
\mathbf{\hat{e}} = \frac{\mathbf{n \times q'}}{\sqrt[]{q'^2 - (\mathbf{n \cdot q'})^2}},
\end{equation}
\begin{equation}
\mathbf{q'} = \mathbf{\hat{B}' + n \times (v \times \hat{B}')} - \frac{\Gamma}{1+\Gamma}(\mathbf{\hat{B}' \cdot v})\mathbf{v}.
\end{equation}
Here prime quantities are measured in the jet frame, $\mathbf{n}$ is the vector to the observer, $\mathbf{\hat{B}}$ is the magnetic field vector and $\mathbf{v}$ is the jet bulk velocity vector.\\

For radiation emitted in a jet zone containing a helical B-field, where
\begin{equation}
\mathbf{\hat{B}'} = \frac{{\rm sin}(\phi_B)\mathbf{\hat{x}'} + {\rm cos}(\phi_B)\mathbf{\hat{y}'} + {\rm tan}(\Psi_B)\mathbf{\hat{z}'}}{\sqrt{1+{\rm tan}^2(\Psi_B)}}
\end{equation}
and $\hat{z}'$ is parallel the jet axis. $\phi_B$ represents the phase angle along the helix and $\Psi_B$ is the pitch angle.
We can see the effects of RPAR on the observed EVPAs for different $\Gamma_{\rm bulk}$ regimes in Figures 2 and 3. The jet shown here is pointed 4$^{\circ}$ off our line of sight in the $\hat{x}$ direction on the plane of the sky. The field  is pitched at $\Psi_B=45^\circ$ to the jet axis.
This pitch angle affects the magnitude of the RPAR effect \citep[as in][]{qian_intrinsic_2004}. RPAR effects are reduced for smaller pitch angles and increased for larger ones; we thus choose $\Psi_B=45^\circ$ for illustrative purposes.
As Figure 3 shows there are three characteristic regimes for the RPAR effects. In the `low $\Gamma$' case RPAR induces a counter-clockwise shift of the EVPA for half a cycle and clockwise for the other half (with the phase controlled by the component of the helical B along the jet axis). This case is represented by the blue dotted line of Figure 2 and the first row of Figure 3. The net effect is that, for a smoothly increasing $\phi_B$ (an EVPA rotating smoothly in the fluid frame), the lab EVPA rotates quickly for half a cycle and slows in the second half, centered perpendicular to the projected jet axis. Thus an intrinsic smooth rotation is seen as a set of `stair-steps' spanning 360$^\circ$.

	For somewhat faster jets, the `high $\Gamma$' case, the RPAR bias (here toward +y) dominates so the EVPA is driven increasingly transverse to the jet. In this regime the observed rotation can switch between clockwise and counter-clockwise rotations for the same field geometry. This is the green dotted line of Figure 2 and the second row of Figure 3. For very large jet bulk motion, the `extreme $\Gamma$' case, the observed EVPA is actually reflected across the jet axis from the initial vector. 
One can understand these behaviors by picturing a relativistic electron emitting synchrotron radiation in the fluid frame within a uniform B-field at arbitrary pitch angle, where the radiation is intermittently directed towards the observer. The electron will emit beamed radiation polarized mostly perpendicular to the B-field. As this emission is boosted to the lab frame, slightly off axis, the radiation initially directed to the side is boosted toward the observer, rotating the PA. For extreme $\Gamma$, the radiation initially at large angle to the line of sight, is boosted toward the observer and begins to dominate the received signal. The EVPA flips across the projected jet axis.
Of course, the bulk of the radiation is boosted into an angle $1/\Gamma$ (red dot-dashed line). Thus to see the `extreme $\Gamma$' case one needs to observe so far off-axis that the observer sees only a tiny portion of the observable jet power. Thus virtually all observers will be in the first two cases, with typical jet alignment resulting in the `Low $\Gamma$' case and only jets viewed well off-axis reaching the `High $\Gamma$' regime. Typical HBL parameters inferred from VLBI observations and SED fits have $\theta_{\rm obs} = 1^{\circ} - 5^{\circ}$, generally in the 'Low $\Gamma$' case. Thus `stair-steps', regular slope variations in extended rotation phases, will often be present, although the amplitude is quite sensitive to field pitch angle and jet parameters.

In Figure 2 one can see that for small $\Gamma$ the observed EVPA is generally driven toward the jet axis by RPAR. Although for the particular pitch angle $\Psi_{\rm B}$ shown there is a range of `Large $\Gamma$' vectors that are driven perpendicular to B, when one averages over all magnetic inclinations the former effect dominates. At large $\Gamma$ the forward boosting drives to the jet axis as well. Thus we see that in Figure 4 that a random set of initial EVPA vectors is driven to the jet axis, except for a range around $1/\theta_{\rm obs}$, which would be seldom observed. As noted in the introduction, a statistical bias for such alignment is indeed observed \citep{jorstad_multifrequency_2006}.

\begin{figure}[t!]
\includegraphics[width=0.97\linewidth, height=6.5cm]{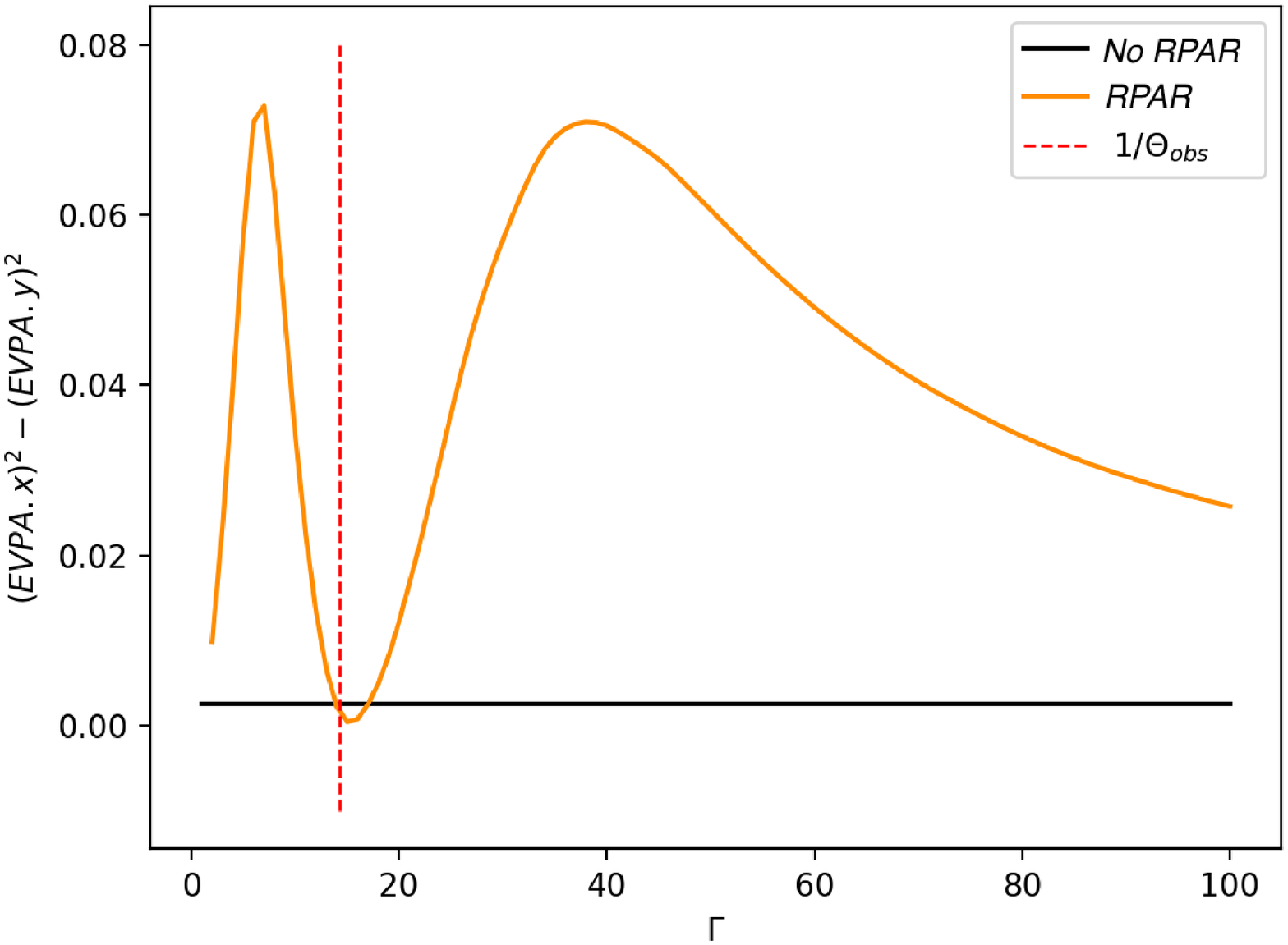}
\centering
\caption{EVPA bias along jet axis against $\Gamma_{bulk}$ for a set of 127 random B-field zones in the jet frame. The bias is calculated as the difference in squared EVPA components parallel ($\hat{x}$) and perpendicular ($\hat{y}$) to the jet axis. The bias is averaged over 500 random realizations. $\theta_{\rm obs} = 4.0^{\circ}$.}
\end{figure}

\section{\label{sec:level1}Geometrical Jet Model}

	The modest, and fluctuating, blazar polarization in the stochastic phases suggests that many zones contribute to the radiation seen with uncorrelated, and varying magnetic field orientation. We thus follow many authors (e.g. Marscher (2014)'s TEMZ model) in assuming a multizone emission region. We attribute this to shock-induced turbulence. While a typical spectral index indicates a saturated polarization level $\Pi_{\rm max} \sim 0.7$, we more commonly observe $\Pi \approx $10\%, indicating $N \sim (\Pi_{\rm max}/\Pi )^2 \sim 50$ uncorrelated zones contribute to the observed emission. At any one time one receives radiation from an angle $\sim 1/\Gamma$ about the line of sight. Thus if the jet opening angle $\theta_{\rm op}$ is $<1/\Gamma$, then this is the full number of zones seen at a given radius; wider jets will have $\sim N$ zones in the observed subset (see Figure 5). With large $\Gamma$ and small viewing angle we in general expect radiation from a single excitation radius (spanning $N$ zones across the jet) to dominate the received radiation at any one time. The detected spectrum will represent a time averaged output of this excited radiating zone as it travels downstream in near-synchrony with its early emission. 

	However, as noted, the prevalence of persistent rotating phases is inconsistent with a simple superposition of random polarization vectors \cite{blinov_robopol:_2016} and the presence of rotations $>\pi$ demands a deterministic underlying structure. Again with high $\Gamma$ and small $\theta_{\rm obs}$ we observe only one jet radial zone at a given time. Thus if there is an underlying helical field passing through the excitation zone (e.g. a standing shock) then only one portion of the helix will be visible to the observer at any time and the polarization will mark the orientation of its field. To implement this, we assume that a fraction of the jet zones are imprinted with this coherent, helically varying, field (shaded zones in Figure 5). Given the modest $\Pi$ during the rotating phase and the similarity of $\sigma_\Pi$ to that of the stochastic phase, the helical coherent field occupies only a small fraction of the observed jet zones. To illustrate strong rotational signals we use $\sim 1/6$ here, although a smaller fraction may be coherent in actual jets.
    
\begin{figure}[t!]
\includegraphics[width=0.97\linewidth, height=5.5cm]{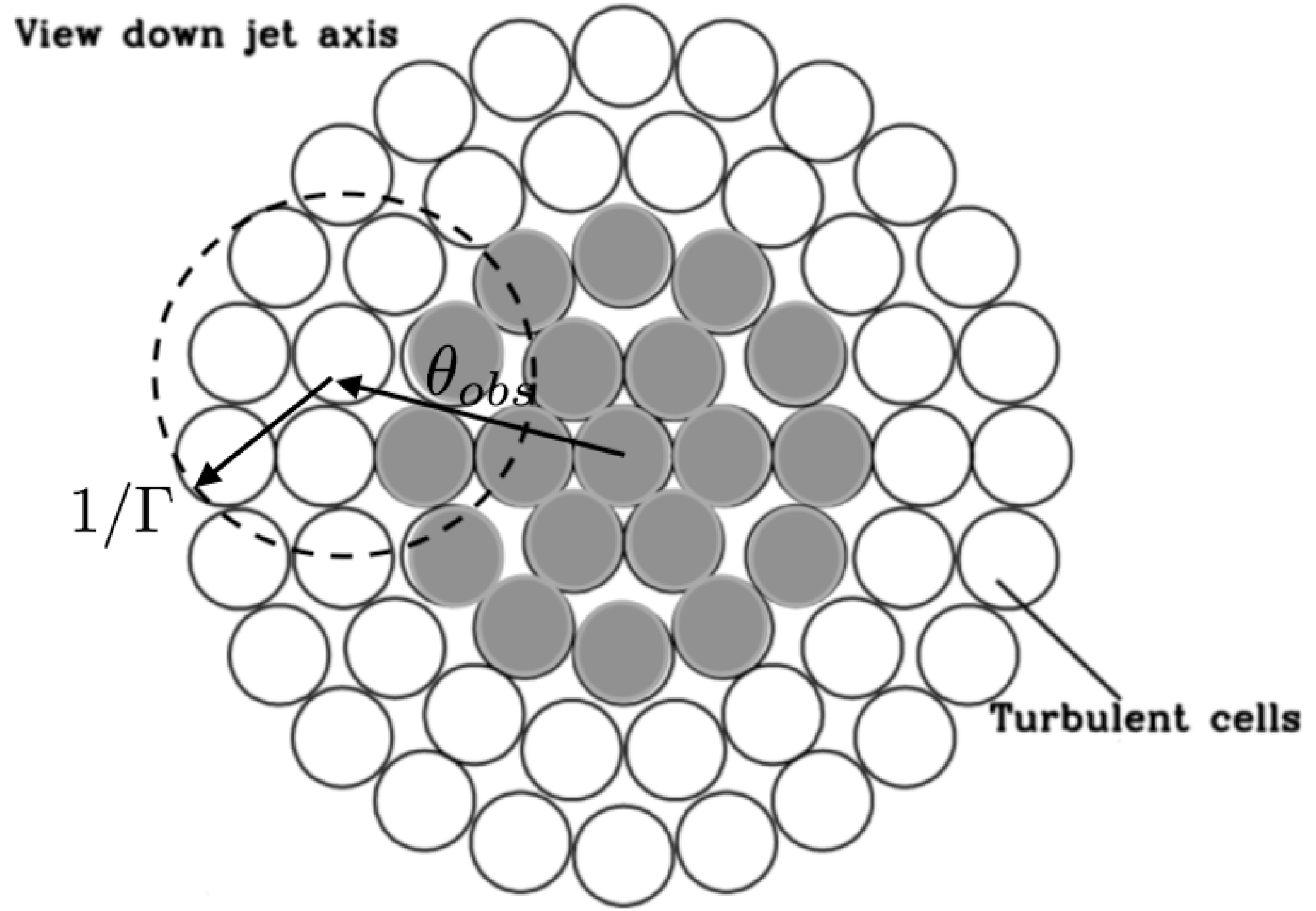}
\centering
\caption{Map of zones in a jet section \cite{marscher_turbulent_2014}. Each zone has a B-field direction, with random initially isotropic orientation during the stochastic phase. During a rotating phase the central zones (shaded) turn helical. The computations typically use 127 zones (6 concentric rings); fewer are shown here for clarity.}
\end{figure}

\begin{figure}[t!]
\includegraphics[width=0.99\linewidth, height=6.5cm]{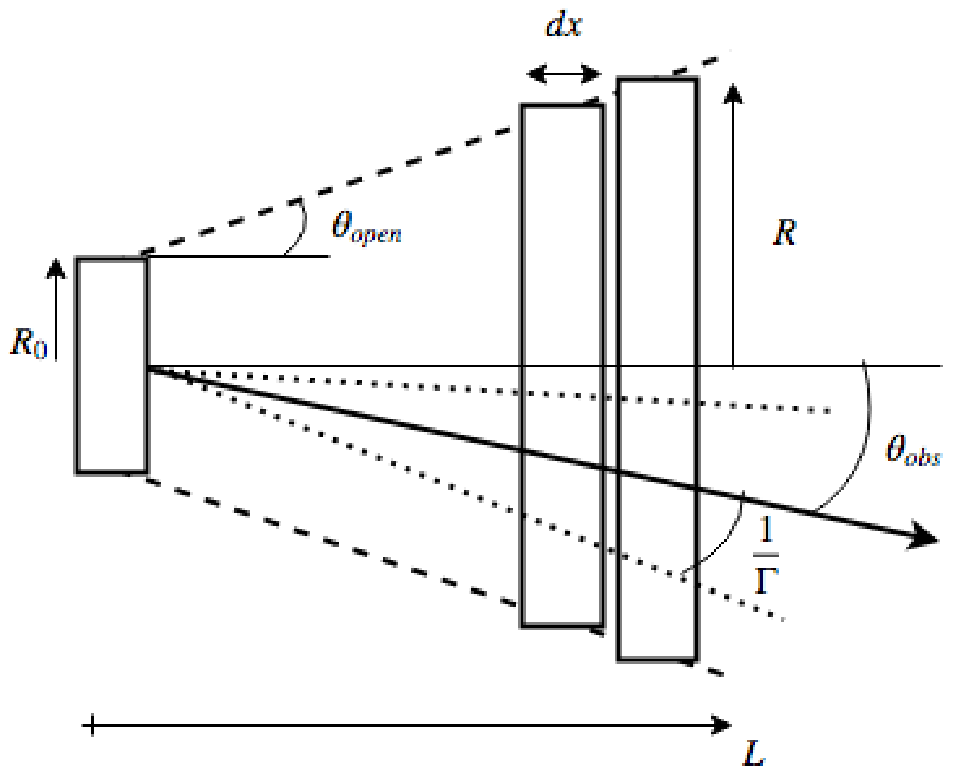}
\centering
\caption{Schematic showing the shape of the `conical' jet model with some of the important parameters labelled, \cite{potter_black_2013}. A `section' of the jet is a $dx$ slice.}
\end{figure}

We expect the jet to be imperfectly collimated, so that its cross section will increase as it propagates. This has two principal effects. First, each zone of the jet has a different $\theta_{\rm obs}$. If one views within $\theta_{\rm op}$ then values from 0 to $\sim {\rm min ({1/\Gamma},\theta_{\rm op})}$ contribute. This averages out some of the RPAR effects. Second, the spreading jet affects the coherent field geometry. Imagine a helical field injected into such a conical jet. Assuming uniform constant $\Gamma$, the distance and time between rotations remains fixed. However the radius of the spiral grows. Hence the pitch angle flattens and the field become more nearly transverse. This leads to evolution in the PA behavior as the jet propagates.

\begin{figure}[h!]
\includegraphics[width=0.97\linewidth, height=6.5cm]{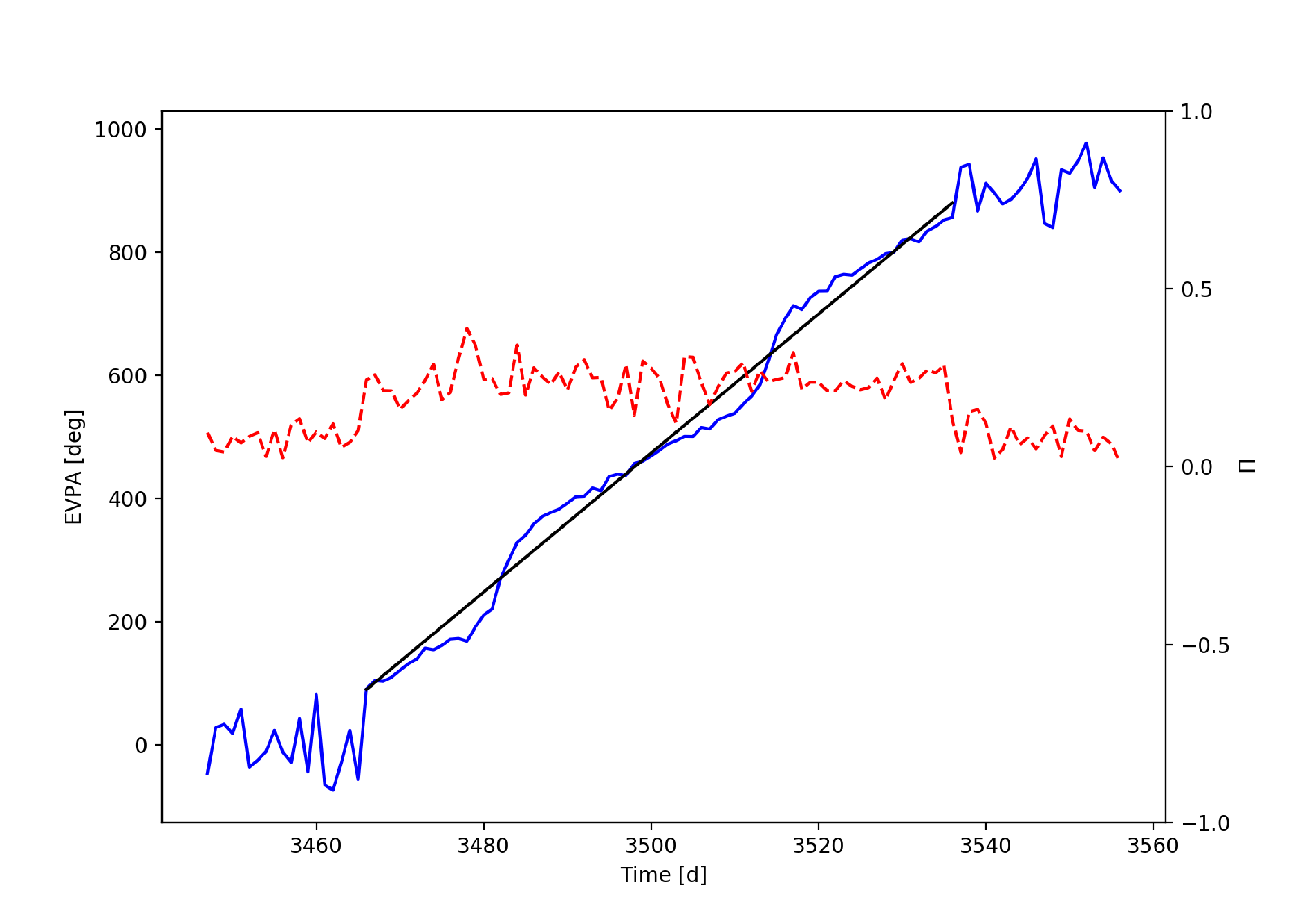}
\centering
\caption{EVPA (blue) and polarization fraction (red) for stochastic and rotating phases in a jet with projection on the plane of the sky along the x axis. We are viewing at $\theta_{\rm obs}=4.0^{\circ}$ with $\Gamma = 5$ and $\theta_{\rm op} = 9.5^{\circ}$. RPAR has been applied. This low $\Gamma$ means the vast majority of zones in the jet contribute. The black line shows the same helical rotation without RPAR or turbulence. The rotation phase spans $770^{\circ}$ with starting phase $\phi = 3\pi/2$.}
\end{figure}

	We now show these geometrical effects in two simulated realizations. Both have low $\Gamma=5$ and opening angle $\theta_{op}=9.5^{\circ}< 1/\Gamma$, and we show a 770$^{\circ}$ rotating bracketed by stochastic phases. RPAR effects and beaming geometry can introduce an especially interesting behavior in extended rotating phases. Especially in the low $\Gamma$ regime, even a smoothly varying helical field produces non-uniform sweep of the observed PA (Figure 3, top line). For extended rotation events we can expect to see this rate modulation as a series of rotation rate steps. Figure 7 shows a simulation of this behavior. This is most prominent when the effective $\theta$ to our line of sight of the helical zones within the $1/\Gamma$ solid angle corresponds with the appropriate $\Gamma$ given by Figures 2 and 3.  Possible examples of such stepped sweeps are seen in PKS 1510-089 \citep{marscher_probing_2010}
and S5 0716+71 \citep{larionov_outburst_2013}.

Of course, in this picture the coherent fields in a portion of the $N$ zones during the rotating phase induce higher polarization. This is not universally seen. In as far as rotations are associated with flare events, turbulence associated with the flare power injection could increase the effective $N$, lowering $\Pi$. Also increased activity during flares may drive higher energy emission closer to the standing shock, where the initially isotropic zones dominate (see below), also lowering $\Pi$. Regardless, as long as the fraction of zones participating in the rotation is not large, the polarization fluctuations $\sigma_\Pi$ will remain substantial and closely resemble those in the non-rotating phase \citep{blinov_robopol:_2016}.

\begin{figure}[h!]
\includegraphics[width=0.97\linewidth, height=6.5cm]{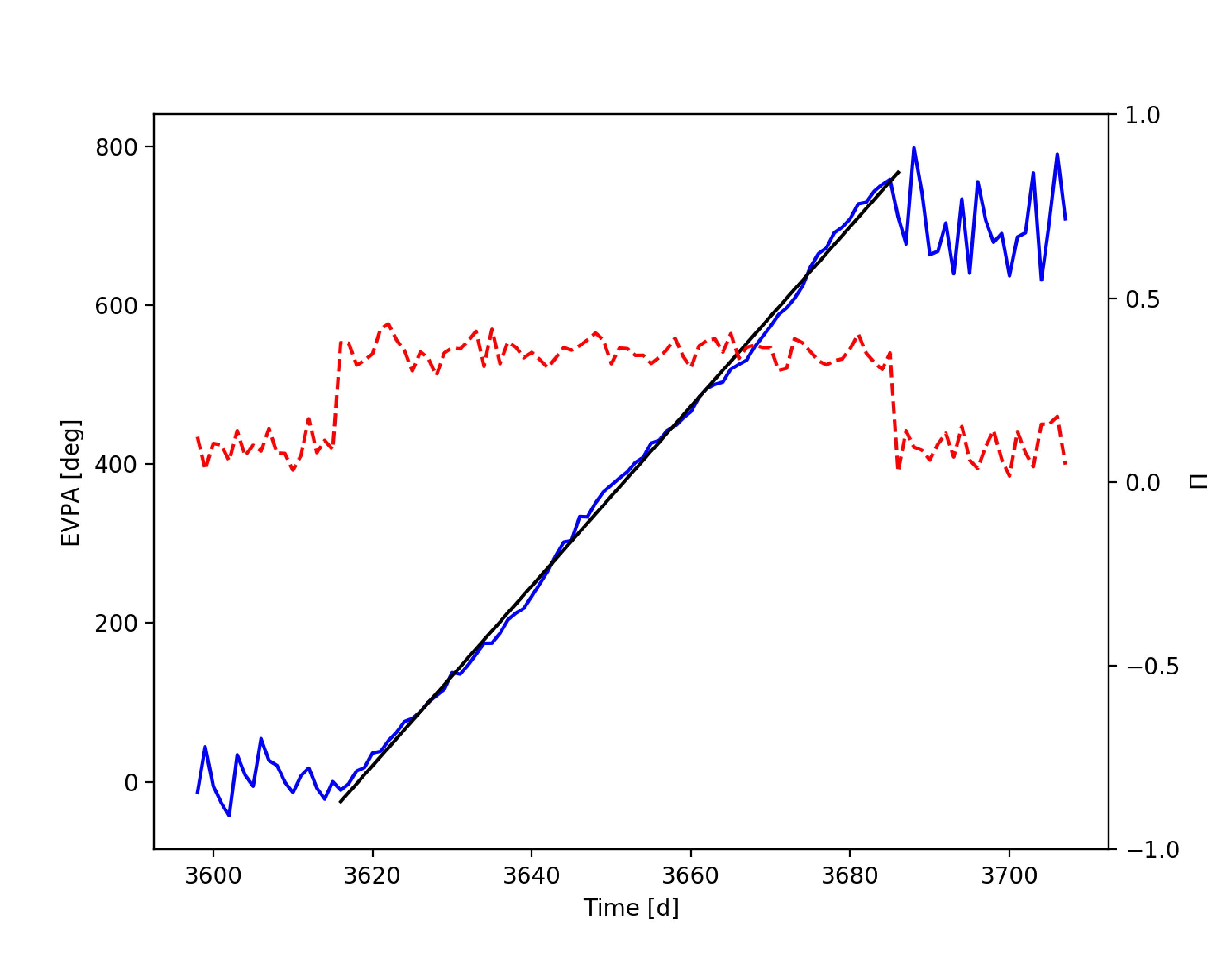}
\centering
\caption{As for Figure 7 for the rotational phases of a conical jet viewed at $\theta_{\rm obs} = 1.5^{\circ}$. This represents a 770$^{\circ}$ rotation with starting phase $\phi=160^{\circ}$. Stepping is less prominent.}
\end{figure}

	Next we show the behavior for a conical jet for the same $\Gamma$ and $\theta_{op}$ but a lower $\theta_{\rm obs}$ in Figure 8. $\Pi$ is larger and fluctuations are smaller since the helical zones now dominate the central position in our line of sight. The PA sweep is steady in the rotation phase now with only small stepping behavior present. A reduction in the effective $\theta$ of the helical zones to our line of sight means that for the same $\Gamma$ we move to the left on Figure 2, bringing us to lower RPAR regime. The stair-stepping effect is much weaker here. Indeed the lack of measurable steps during the rotation phase of J1751+0939 (Figure 1) places some limits on $\theta_{\rm obs}$ and $\Gamma$. This source has a VLBI estimated $\beta_\perp = 7.9\pm 0.8$ \citep{lister_mojave._2013} and we find that $\theta_{\rm obs} < 2^{\circ}$ and $\Gamma > 11$ when pitch angle $\Psi_{\rm B}$ is near $0^{\circ}$. Steeper pitch angles place more stringent constraints, e.g. $\Psi_{\rm B} = 25^{\circ}$ implies $\theta_{\rm obs} \leq 0.8^{\circ}$ and $\Gamma \geq 16$.
    
\section{\label{sec:level1}Radiation Model}

	Our radiation model is inspired by \cite{potter_black_2013} and uses the basic jet setup shown in Figure 6. We take the jet to be composed of an electron-positron plasma (hereafter electron). Under the assumption of equipartition between electron energy and magnetic field energy, the radius and electron population are initialized at the base of the jet from the input parameters: the length of the jet $L_0$, the total jet power $W_j$, the bulk Lorentz factor $\Gamma$, the magnetic field strength at the base $B_0$, the maximum initial electron energy $E_{max}$, the electron power law index $\alpha$, the jet observation angle $\theta_{\rm obs}$ and the jet opening angle $\theta_{op}$. Furthermore, it is assumed that the electrons and magnetic field are homogeneously distributed throughout each jet section and that the magnetic field energy is conserved.
    
	The initial electron population is assumed to have a power law energy distribution with an exponential cutoff \citep{bregman_diffusive_1985} and is represented by a discrete set of energy bins.
\begin{equation}
N(E_e) = AE_e^{-\alpha}e^{\frac{-E_e}{E_{max}}}.
\end{equation}
For each jet zone we divided into sections of length $dx$ (see Fig. 6). In a particular section, the synchrotron emission from each electron energy bin is calculated using the B field in that section. The synchrotron losses then determine the evolution in the electron energy bin populations for the next section. Furthermore, since we are viewing down the jet, the self-absorption opacities for each jet section are calculated and applied to the spectrum - these depend on $\theta_{\rm obs}$ and synchrotron photon energy. This is a quiescent jet model, without flux variability in time; only the B-field directions in the zones change for the purpose of polarization modeling.

	To find the polarization in the conical jet, we calculate the electrons' synchrotron power per unit frequency perpendicular and parallel to the B-field projection on the plane of the sky for each zone in each section. These are \citep{rybicki_radiative_1979}:
\begin{equation}
P_{\perp}(\omega) = \frac{\sqrt{3}q^3B{\rm sin}\alpha}{4\pi mc^2}\big[F(x) + G(x)\big],
\end{equation}
\begin{equation}
P_{\parallel}(\omega) = \frac{\sqrt{3}q^3B{\rm sin}\alpha}{4\pi mc^2}\big[F(x) - G(x)\big].
\end{equation}
Here $F(x)$ and $G(x)$ are the modified Bessel functions \citep[e.g.][]{longair_high_2011}. This radiation is then subject to RPAR and the intensity is Doppler boosted by $\delta^4$. We next sum, using Stokes' parameters, the contribution of all zones and sections to obtain the total powers in a coordinate system aligned with the projected jet axis on the plane of the sky. These then provide the polarization fraction:
\begin{equation}
\Pi(\omega) = \frac{P_{\perp}(\omega) - P_{\parallel}(\omega)}{P_{\perp}(\omega) + P_{\parallel}(\omega)}.
\end{equation}
The projected net EVPA (relative to the jet axis) can also be referenced to an absolute angle.

	Note that in this model the electron population evolves (cools) along the jet. For a conical jet we assume that the field also evolves, becoming increasingly transverse as the jet expands. This means that we expect measurable differences in $\Pi$ and PA as a function of observed frequency. Since the high energy electrons cool most quickly, for appropriate parameters their radiation may come only from the base of the jet. Lower energies come from a large range of jet radii and thus, for a conical jet, a more transverse field structure. Thus we expect an energy-dependent shift in the observed polarization properties. Note also that, for  a given number of emission zones $N$, the conical jet model will have a higher polarization than the simple cylindrical jet case. This is both because the jet divergence means that the received radiation is dominated by zones close to the line of sight and because down-stream fields are increasingly transverse and hence, even for the stochastic case, increasingly coherent. In particular, at low (radio) energies one expects EVPA increasingly aligned with the projected jet axis (Figure 9). This is indeed observed.
 
The change in the effective B field pitch angle as one moves along the jet also introduces energy-dependent shifts in the EVPA at a given phase. From the underlying geometry, with small $\theta_{\rm obs}$ and conical jets, the effects are relatively subtle, since with the jet viewed nearly end-on, even dramatic pitch angle changes make modest change to the projection on the sky. However, RPAR effects make the observed PA sensitive to the full polarization vector and greatly enhance the sensitivity to the magnetic field inclination, even for nearly aligned jets.

Interestingly, this can introduce a rotation-phase dependent modulation of amplitude ${\bar A}$ in the degree of polarization $\Pi$, which is especially strong for the low energy synchrotron emission from larger distances in the spread jet (Figure 10). It will be interesting to see if this pattern can be recovered from monitoring observations at high (core dominated) radio frequencies.

\begin{figure}[h]
\includegraphics[width=0.97\linewidth, height=6.5cm]{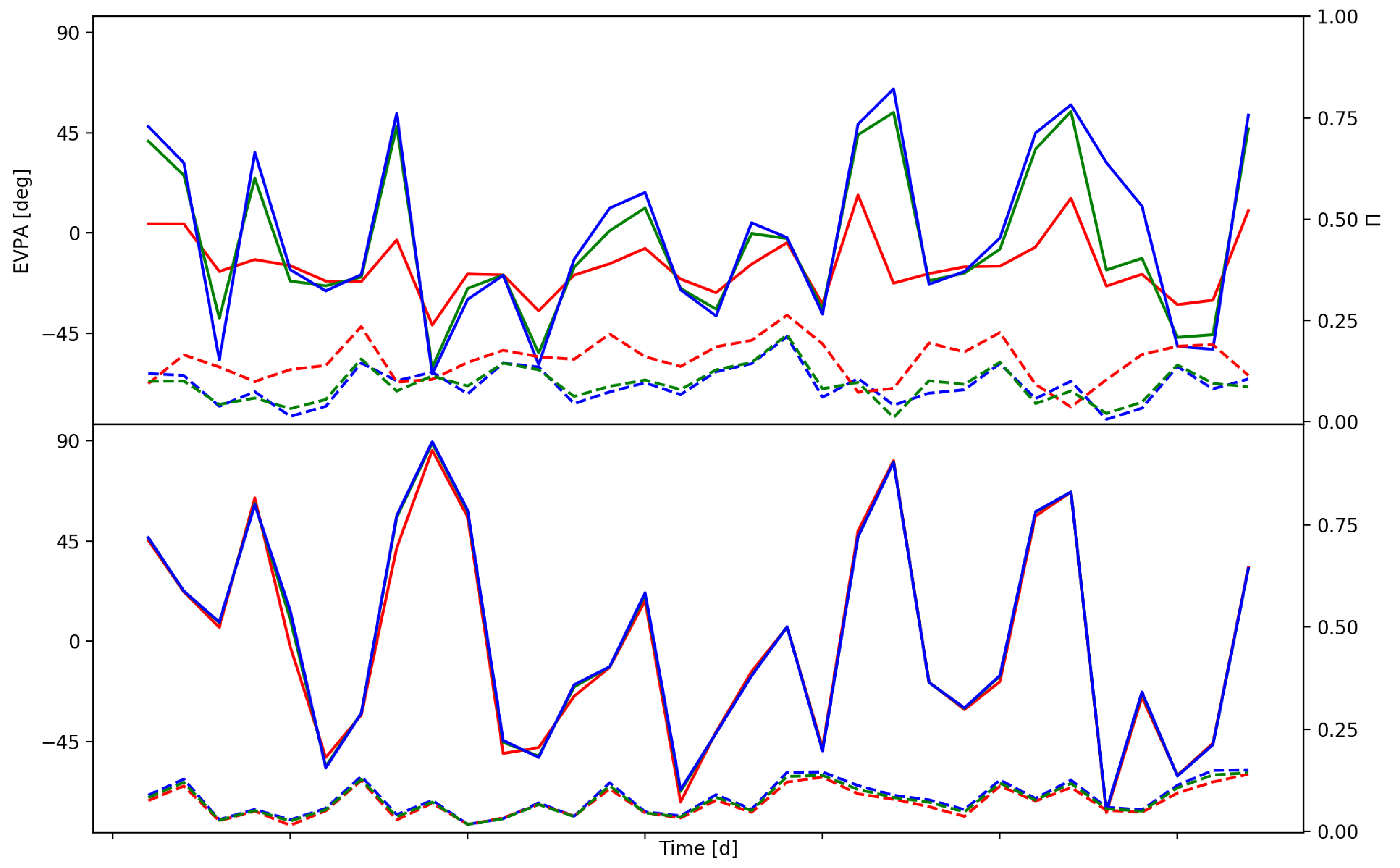}
\centering
\caption{Energy variation for EVPA and $\Pi$ during a non-rotating (stochastic) phase. Colors denote low (Radio, red), mid-range (Optical, green) and near cut-off (e.g. X-ray, blue) synchrotron bands. This is for a jet with $\theta_{\rm obs} = 1.5^{\circ}$, $\Gamma = 5$ and $\theta_{op} = 9.5^{\circ}$. Both panels use the random seed for the generated B-fields. The top panel shows a RPAR affected jet: substantial energy dependence is seen, with low energy (radio) PA better aligned with the parent jet. Low energy fluctuations are also somewhat smaller. In the lower panel RPAR effects are ignored and the behavior is essentially achromatic with lower overall polarization fraction.}
\end{figure}

\begin{figure}[h]
\includegraphics[width=0.97\linewidth, height=6.5cm]{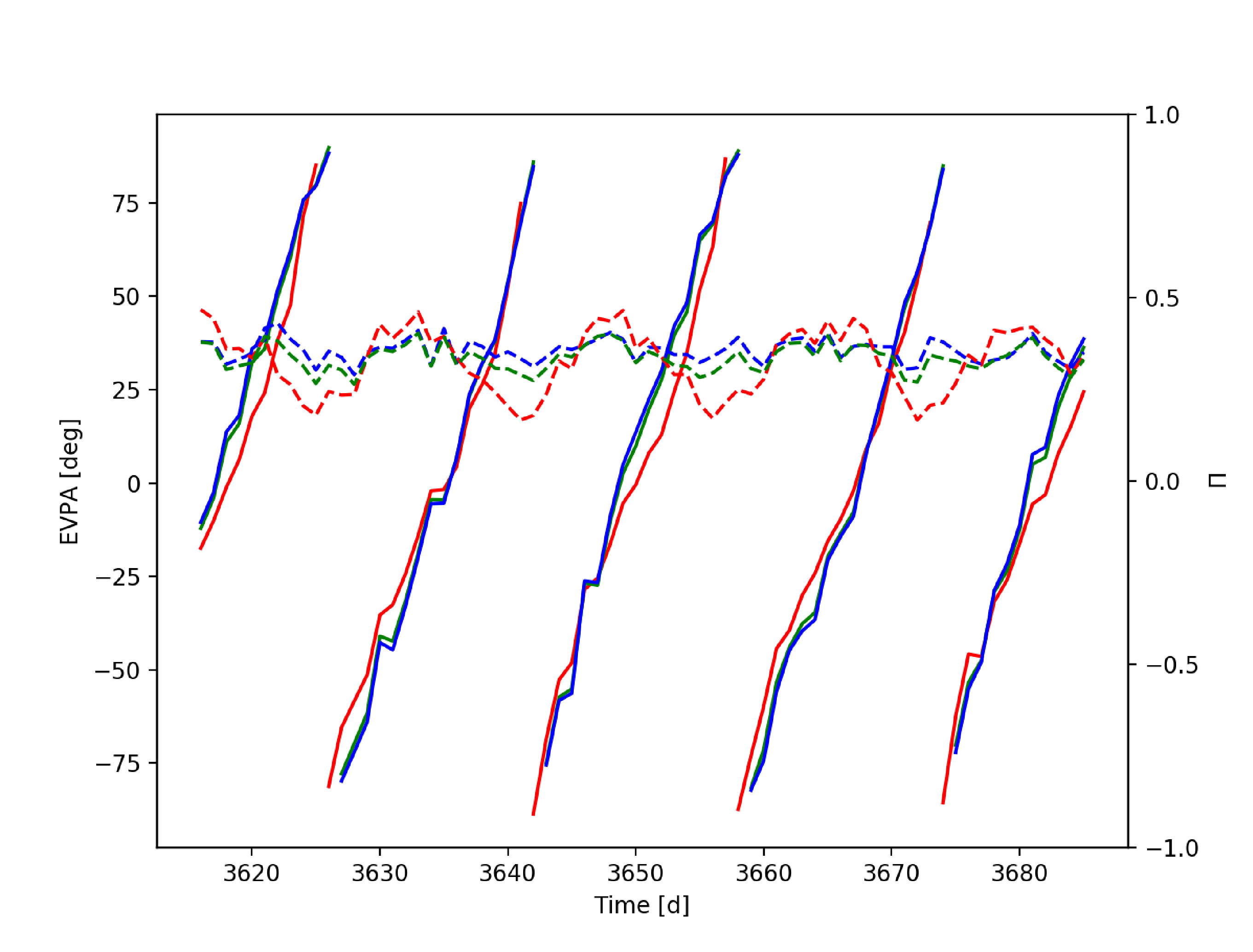}
\centering
\caption{Energy dependence during a rotating phase (the $770^{\circ}$ rotation shown in Figure 8), as for Figure 9. Strong sinusoidal variation of the radio polarization fraction occurs as the rotating helical zone EVPA aligns parallel and perpendicular to the jet axis. Since the random zones' EVPA are increasingly aligned with this axis as one progresses along the jet (i.e. observes at lower frequency; see top panel, Figure 9) the total $\Pi$ has a sinusoidal modulation. The strength of the modulation depends on the ratio of helical zones to random zones.}
\end{figure}
\begin{table}[h!]
\centering
\begin{tabular}{l c c c c c c}
Fig &
$\bar{\Pi}_{R}$ & $\bar{\sigma}_{R}$ & $\bar{A}_{R}$ & $\bar{\Pi}_{X}$ & $\bar{\sigma}_{X}$ & $\bar{A}_{X}$ \\ 
 \hline\hline
9 & $0.15$ & $0.051$ & $---$ & $0.09$ & $0.045$ & $---$ \\ 
10 & $0.32$ & $0.031$ & $0.11$ & $0.36$ & $0.029$ & $0.016$ \\ 
 \hline
\end{tabular}
\caption{Radio and X-ray polarization fractions for Figures 9 (top panel) and 10. For the Figure 9 row $\bar{\sigma}$ describes the fluctuations about the mean $\bar{\Pi}$, while for Figure 10 it 
represents the uncertainty in the amplitude ${\bar A}$ of the best-fit sinusoid.}
\end{table}

By taking the electron population to be homogeneously distributed across the jet zones, our model assumes the main source of energy dependence in polarization to originate from the B-field change along the jet linked with RPAR. However, having higher energy electrons relegated to fewer zones as in \cite{marscher_rapid_2010} or \cite{angelakis_robopol:_2016} has been invoked to explain the blazar sequence polarization trends described in the studies of \cite{itoh_systematic_2016} and \cite{angelakis_robopol:_2016} which suggest LBLs have higher polarization fraction (up to 40\%) and variability than HBLs (up to 10\%) in the optical. Indeed if the X-ray polarization fraction is observed to be much greater than the optical in HBL sources, one could plausibly extend this model by having X-ray synchrotron emission from fewer zones as above. \\ 
In this paper we treat only polarization of the synchrotron emission, so this model describes up to the peak $\nu_{\rm sy}$. This is in the X-ray band for HBL, but typically in the IR/optical band for other blazars. Treatment of Inverse-Compton regime polarization is covered in \cite{zhang_polarization_2016}. In general IC components should have a lower polarization fraction, including when SSC dominates. We expect this will decrease the observed $\Pi$ in our model for a given number of radiation zones $N$. When external photon fields dominate, the observed $\Pi$ will be even smaller. 

\section{\label{sec:level1}Application to MRk 501}
	One motivation for this study is the new prospect of measuring blazar X-ray polarization with {\it IXPE} \citep{weisskopf_imaging_2016} or other upcoming missions. To date only a handful of X-ray polarization measurements have been made (OSO-8, PoGo+, X-Calibur) mostly of the Crab Nebula with no blazar measurements as of yet \citep{kislat_optimization_2018}. However new facilities should provide a number of good polarization measurements, making an evaluation of energy dependence timely.
For many LBLs, thermal emission is important in the optical band and synchrotron emission does not dominate. Also, the radio emission is often dominated by larger scale jet flux far from the acceleration zone. So intraband comparisons of HBLs are especially interesting since the optical and even the X-ray can come from the synchrotron peak, allowing multiband comparisons to probe the RPAR and jet geometry effects described above. 
    
    We thus illustrate the various polarization phenomena with simulations of an HBL source,  Mrk 501. This blazar displays substantial optical PA variability, including EVPA rotation and can be well measured by IXPE in a few day's exposure. Mrk 501's Doppler factor $\delta$ has been estimated as $\sim 6-22$, leaving much leeway in choosing the $\Gamma$, $\theta_{\rm obs}$ and $\theta_{\rm op}$. As shown in \textsection2, different $\Gamma$ lead to different RPAR behaviors, so we have modeled for two values consistent with allowed $\delta$ range, adjusting jet parameters to fit the overall SED (bottom panel of Figure 11). Table 2 shows the selected fit parameters. 

\begin{figure}[h]
\includegraphics[width=0.97\linewidth, height=6.5cm]{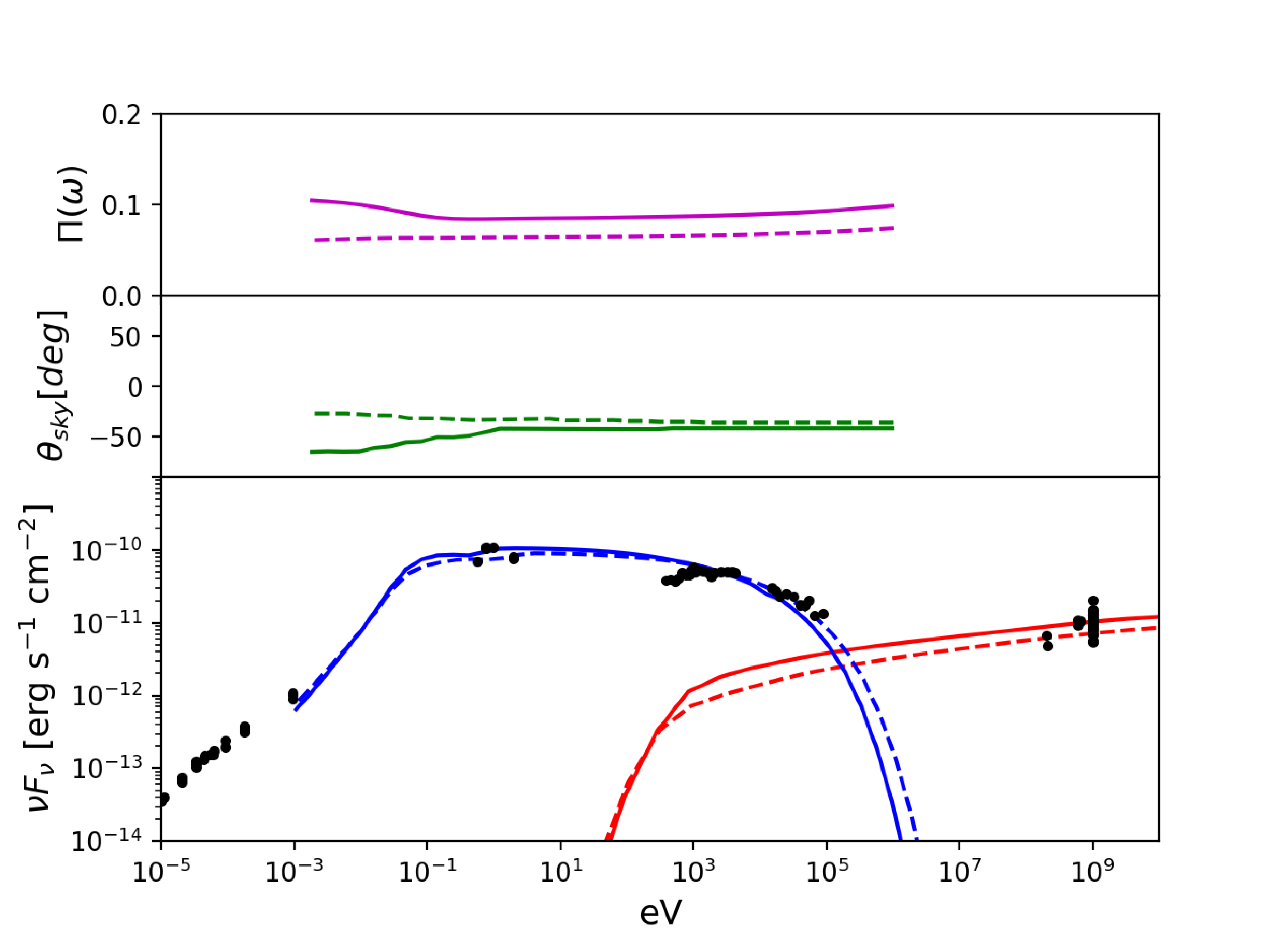}
\centering
\caption{A fit to Mrk 501's SED (bottom panel, approximately contemporaneous fluxes drawn from the ASDC compilation for 2010; {\tt https://tools.asdc.asi.it/SED/}) using the radiative model. The top two panels show the energy dependence of $\Pi$ and EVPA averaged over 200 iterations for the stochastic phase, with solid line for a low $\delta$ solution and a dashed line for high $\delta$. The observed EVPA is expected to fluctuate about these mean values during stochastic phases. The jet projection on the plane of the sky is $\theta_{\rm sky} =-40^{\circ}$, as observed for Mrk 501.}
\end{figure}
\begin{table}[h!]
\centering
\begin{tabular}{c c c c c c c} 
$\delta$ & $B_0[G]$ & $\gamma_{\rm max}$ & $\alpha$ & $\theta_{\rm op}[^{\circ}]$ & $\theta_{\rm obs}[^{\circ}]$ & $\Gamma$ \\ 
 \hline\hline
$6.6$ & $40$ & $4.9*10^{4}$ & $1.95$ & $13.0$ & $4.0$ & $7.5$ \\ 
 \hline
$17.4$ & $30$ & $6.3*10^{4}$ & $1.95$ & $3.9$ & $3.3$ & $17.5$ \\ 
 \hline
\end{tabular}
\caption{Parameters used for the Mrk 501 fits of Figure 11. The top and bottom rows are for the solid and dashed fits respectively. Both fits use $W_j = 2*10^{37}W$, $L_j = 5*10^{20}m$ and $\gamma_{\rm min}=10$.}
\end{table}

Both sets of parameters fit the SED reasonably well while producing different polarization behavior. The solid line set provides a higher average polarization fraction across all bands as slightly fewer B-field zones lie within its $1/\Gamma$ range (Figure 5). Also, for this model, rotating phases (not shown) will produce strongly `stepped modulation', due to the proximity of the helical zones to the line of sight. In contrast, the dashed line set does not produce EVPA rotations since the helical zones are located on the periphery of its more restricted $1/\Gamma$ range. Note that in this picture, we can predict that low $\Gamma$ jets are, in general, more likely to produce rotation events. Finally, the difference in the low energy polarization behavior provides additional observables that are sensitive to $\theta_{\rm obs}$ and $\theta_{\rm op}$.

\section{\label{sec:level1}Conclusions}

We have explored a simple geometrical model of a conical blazar jet with multiple emission zones across the interior. By introducing a coherently rotating helical field in a subset of these emission zones and by noting that the jet magnetic fields can become increasingly transverse as the jet expands, we have been able to mimic a variety of observed blazar PA behavior. Our study treats the often neglected effect of the jet boost on the observed EVPA (RPAR) which provides interesting effects on the observed PA behavior in some regimes. Finally, by computing with a simple emission model in this jet geometry we have seen that the energy dependence of some polarization observables can also display useful dependence on the model parameters. 

	Although many of our ingredients have been considered in past studies, by combining these into a single model, we have been able to reproduce a large fraction of the observed EVPA phenomena. Moreover, the model makes interesting predictions for the (modest) energy-dependence of EVPA across the synchrotron peak of the blazar emission. With the possibility of X-ray polarization measurements in the near future, the model allows for some useful comparison with observed data sets. Novel dependence of the EVPA observables on $\Gamma$ and $\theta_{\rm obs}$ offer new ways of constraining these important parameters. 
    
Of course extensions are needed: for most blazars IC is relevant in X-ray 
emission, so we should add simulation of this more weakly polarized emission to the model. But the interesting patterns in $\Pi$ and PA introduced by the jet expansion and the relativistic boost provide a range of observables that can be sought in extensive polarization monitoring programs. These patterns can be useful in constraining jet geometry and, eventually, in guiding detailed RMHD modeling that will follow the shocks and acceleration giving rise to the energetic electron populations responsible for the observed (polarized) blazar emission. 
    
\bigskip
	We thank I. Liodakis for discussions on the state of optical polarization measurements. This work was supported in part by grant NNM17AA26C.
    
\bibliographystyle{apj}
\bibliography{Zotero.bib}

\begin{thebibliography}{}

\bibitem[\protect\citeauthoryear{Abdo et~al.}{Abdo
  et~al.}{2010}]{abdo_spectral_2010}
Abdo, A.~A., et~al. 2010, ApJ, 716, 30

\bibitem[\protect\citeauthoryear{Angelakis et~al.}{Angelakis
  et~al.}{2016}]{angelakis_robopol:_2016}
Angelakis, E., et~al. 2016, MNRAS, 463, 3365, arXiv: 1609.00640

\bibitem[\protect\citeauthoryear{Blandford \& Koenigl}{Blandford \&
  Koenigl}{1979}]{blandford_relativistic_1979}
Blandford, R.~D.,  \& Koenigl, A. 1979, ApJ, 232, 34

\bibitem[\protect\citeauthoryear{Blandford \& Znajek}{Blandford \&
  Znajek}{1977}]{blandford_electromagnetic_1977}
Blandford, R.~D.,  \& Znajek, R.~L. 1977, MNRAS, 179, 433

\bibitem[\protect\citeauthoryear{Blinov et~al.}{Blinov
  et~al.}{2018}]{blinov_robopol:_2018}
Blinov, D., et~al. 2018, MNRAS, 474, 1296

\bibitem[\protect\citeauthoryear{Blinov et~al.}{Blinov
  et~al.}{2015}]{blinov_robopol:_2015}
Blinov, D., et~al. 2015, MNRAS, 453, 1669

\bibitem[\protect\citeauthoryear{Blinov et~al.}{Blinov
  et~al.}{2016}]{blinov_robopol:_2016}
Blinov, D., et~al. 2016, MNRAS, 457, 2252

\bibitem[\protect\citeauthoryear{Bregman}{Bregman}{1985}]{bregman_diffusive_1985}
Bregman, J.~N. 1985, ApJ, 288, 32

\bibitem[\protect\citeauthoryear{D'arcangelo et~al.}{D'arcangelo
  et~al.}{2009}]{darcangelo_synchronous_2009}
D'arcangelo, F.~D., et~al. 2009, ApJ, 697, 985

\bibitem[\protect\citeauthoryear{Gabuzda, Murray, \& Cronin}{Gabuzda
  et~al.}{2004}]{gabuzda_helical_2004}
Gabuzda, D.~C., Murray, Ã.,  \& Cronin, P. 2004, MNRAS, 351, L89

\bibitem[\protect\citeauthoryear{Hughes, Aller, \& Aller}{Hughes
  et~al.}{1989}]{hughes_synchrotron_1989}
Hughes, P.~A., Aller, H.~D.,  \& Aller, M.~F. 1989, ApJ, 341, 68

\bibitem[\protect\citeauthoryear{Itoh et~al.}{Itoh
  et~al.}{2016}]{itoh_systematic_2016}
Itoh, R., et~al. 2016, ApJ, 833, 77, arXiv: 1610.04313

\bibitem[\protect\citeauthoryear{Jorstad et~al.}{Jorstad
  et~al.}{2006}]{jorstad_multifrequency_2006}
Jorstad, S., et~al. 2006, ChJP, 6, 247

\bibitem[\protect\citeauthoryear{Kislat et~al.}{Kislat
  et~al.}{2018}]{kislat_optimization_2018}
Kislat, F., Abarr, Q., Beheshtipour, B., Geronimo, G.~D., Dowkontt, P., Tang,
  J.,  \& Krawczynski, H. 2018, JATIS, 4, 011004

\bibitem[\protect\citeauthoryear{Larionov et~al.}{Larionov
  et~al.}{2013}]{larionov_outburst_2013}
Larionov, V.~M., et~al. 2013, ApJ, 768, 40

\bibitem[\protect\citeauthoryear{Lister et~al.}{Lister
  et~al.}{2013}]{lister_mojave._2013}
Lister, M.~L., et~al. 2013, AJ, 146, 120

\bibitem[\protect\citeauthoryear{Longair}{Longair}{2011}]{longair_high_2011}
Longair, M.~S. 2011, High {Energy} {Astrophysics} (Cambridge University Press),
  Google-Books-ID: KGe3FVbDNk4C

\bibitem[\protect\citeauthoryear{Lynch et~al.}{Lynch
  et~al.}{2018}]{lynch_green_2018}
Lynch, R.~S., et~al. 2018, ApJ, 859, 93

\bibitem[\protect\citeauthoryear{Lyutikov \& Kravchenko}{Lyutikov \&
  Kravchenko}{2017}]{lyutikov_polarization_2017}
Lyutikov, M.,  \& Kravchenko, E. 2017, MNRAS, 467, 3876, arXiv: 1702.02354

\bibitem[\protect\citeauthoryear{Lyutikov, Pariev, \& Blandford}{Lyutikov
  et~al.}{2003}]{lyutikov_polarization_2003}
Lyutikov, M., Pariev, V.~I.,  \& Blandford, R. 2003, ApJ, 597, 998, arXiv:
  astro-ph/0305410

\bibitem[\protect\citeauthoryear{Maraschi, Ghisellini, \& Celotti}{Maraschi
  et~al.}{1992}]{maraschi_jet_1992}
Maraschi, L., Ghisellini, G.,  \& Celotti, A. 1992, ApJ, 397, L5

\bibitem[\protect\citeauthoryear{Marscher}{Marscher}{2014}]{marscher_turbulent_2014}
Marscher, A.~P. 2014, ApJ, 780, 87

\bibitem[\protect\citeauthoryear{Marscher \& Jorstad}{Marscher \&
  Jorstad}{2010}]{marscher_rapid_2010}
Marscher, A.~P.,  \& Jorstad, S.~G. 2010, arXiv:1005.5551 [astro-ph], arXiv:
  1005.5551

\bibitem[\protect\citeauthoryear{Marscher et~al.}{Marscher
  et~al.}{2010}]{marscher_probing_2010}
Marscher, A.~P., et~al. 2010, ApJ, 710, L126

\bibitem[\protect\citeauthoryear{Nalewajko}{Nalewajko}{2017}]{nalewajko_model_2017}
Nalewajko, K. 2017, Galaxies, 5, 64, arXiv: 1711.00899

\bibitem[\protect\citeauthoryear{Potter \& Cotter}{Potter \&
  Cotter}{2013}]{potter_black_2013}
Potter, W.~J.,  \& Cotter, G. 2013, Ph.D. thesis, Oxford University, UK

\bibitem[\protect\citeauthoryear{Qian \& Zhang}{Qian \&
  Zhang}{2004}]{qian_intrinsic_2004}
Qian, S.-J.,  \& Zhang, X.-Z. 2004, Chin. J. Astron. Astrophys., 4, 37

\bibitem[\protect\citeauthoryear{Rybicki \& Lightman}{Rybicki \&
  Lightman}{1979}]{rybicki_radiative_1979}
Rybicki, G.~B.,  \& Lightman, A.~P. 1979, Radiative {Processes} in
  {Astrophysics} (John Wiley \& Sons), Google-Books-ID: LtdEjNABMlsC

\bibitem[\protect\citeauthoryear{Urry \& Padovani}{Urry \&
  Padovani}{1995}]{urry_unified_1995}
Urry, C.~M.,  \& Padovani, P. 1995, Publ Astron Soc Pac, 107, 803, arXiv:
  astro-ph/9506063

\bibitem[\protect\citeauthoryear{Weisskopf et~al.}{Weisskopf
  et~al.}{2016}]{weisskopf_imaging_2016}
Weisskopf, M.~C., et~al. 2016, Results Phys, 6, 1179

\bibitem[\protect\citeauthoryear{Zhang et~al.}{Zhang
  et~al.}{2015}]{zhang_polarization_2015}
Zhang, H., Chen, X., Böttcher, M., Guo, F.,  \& Li, H. 2015, ApJ, 804, 58

\bibitem[\protect\citeauthoryear{Zhang et~al.}{Zhang
  et~al.}{2016}]{zhang_polarization_2016}
Zhang, H., Deng, W., Li, H.,  \& Böttcher, M. 2016, ApJ, 817, 63

\end{thebibliography}
\end{document}